\documentclass[twocolumn,times,tighten]{aastex63}

\usepackage{amsmath,amstext}
\usepackage[english]{babel}
\usepackage[utf8x]{inputenc}
\usepackage{textgreek}

\newcommand{\myedit}[1]{{#1}}

\newcommand{\teff}{$T_\mathrm{eff}$}
\newcommand{\logg}{$\log g$}
\newcommand{\feh}{[Fe/H]}
\newcommand{\micro}{$\xi_\mathrm{micro}$}
\newcommand{\sife}{[Si/Fe]}
\newcommand{\afe}{[\textalpha/Fe]}
\newcommand{\wm}{{\color{white}$-$}}
\newcommand{\wo}{{\color{white}$1$}}

\definecolor{Green}{rgb}{0.15,0.45,0.25}

\shorttitle{Chemistry of the Nuclear Star Cluster}
\shortauthors{Thorsbro et al.}

\begin{document}



\title{Detailed abundances in the Galactic center:
Evidence of a metal-rich alpha-enhanced stellar population}

\author[0000-0002-5633-4400]{B. Thorsbro}
\affil{Lund Observatory, Department of Astronomy and Theoretical Physics, Lund University, Box 43, SE-22100 Lund, Sweden}
\email{thorsbro@astro.lu.se}

\author[0000-0001-6294-3790]{N. Ryde}
\affil{Lund Observatory, Department of Astronomy and Theoretical Physics, Lund University, Box 43, SE-22100 Lund, Sweden}

\author[0000-0003-0427-8387]{R. M. Rich}
\affil{Department of Physics and Astronomy, UCLA, 430 Portola Plaza, Box 951547, Los Angeles, CA 90095-1547, USA}

\author[0000-0002-6590-1657]{M. Schultheis}
\affil{Observatoire de la C\^ote d'Azur, CNRS UMR 7293, BP4229, Laboratoire Lagrange, F-06304 Nice Cedex 4, France}

\author[0000-0001-5073-2267]{F. Renaud}
\affil{Lund Observatory, Department of Astronomy and Theoretical Physics, Lund University, Box 43, SE-22100 Lund, Sweden}

\author[0000-0001-9715-5727]{E. Spitoni}
\affil{Stellar Astrophysics Centre, Department of Physics and Astronomy, Aarhus University, Ny Munkegade 120, DK-8000 Aarhus C, Denmark}

\author[0000-0002-3122-300X]{T. K. Fritz}
\affil{Instituto de Astrofísica de Canarias, Calle Via Láctea s/n, E-38206 La Laguna, Tenerife, Spain}
\affil{Universidad de La Laguna. Avda. Astrofísico Fco. Sánchez, La Laguna, Tenerife, Spain}

\author[0000-0002-2386-9142]{A. Mastrobuono-Battisti}
\affil{Lund Observatory, Department of Astronomy and Theoretical Physics, Lund University, Box 43, SE-22100 Lund, Sweden}

\author[0000-0002-6040-5849]{L. Origlia}
\affil{INAF---OAS Osservatorio di Astrofisica e Scienza dello Spazio di Bologna, Via Gobetti 93/3,I-40129 Bologna, Italy}

\author[0000-0002-6040-5849]{F. Matteucci}
\affil{Dipartimento di Fisica, Sezione di Astronomia, Universit\`a di Trieste, via G.B. Tiepolo 11, I-34131, Trieste, Italy}
\affil{I.N.A.F. Osservatorio Astronomico di Trieste, via G.B. Tiepolo 11, I-34131, Trieste, Italy}
\affil{I.N.F.N. Sezione di Trieste, via Valerio 2, 34134 Trieste, Italy}

\author[0000-0001-5404-797X]{R. Schödel}
\affil{Instituto de Astrofísica de Andalucía (CSIC), Glorieta de la Astronomía s/n, 18008 Granada, Spain}

\begin{abstract}
We present a detailed study of the composition of 20 M giants in the Galactic center with 15 of them confirmed to be in the Nuclear Star Cluster. As a control sample we have also observed 7 M giants in the Milky Way Disk with similar stellar parameters. All 27 stars are observed using the NIRSPEC spectograph on the KECK II telescope in the K-band at a resolving power of $R=23,000$. 
We report the first silicon abundance trends versus [Fe/H] for stars in the Galactic center. While finding a disk/bulge like trend at subsolar metallicities, we find that [Si/Fe] is enhanced at supersolar metallicities.  We speculate on possible enrichment scenarios to explain such a trend. However, the sample size is modest and the result needs to be confirmed by additional measurements of \myedit{silicon and} other \textalpha-elements. We also derive a new distribution of [Fe/H] and find the most metal rich stars at [Fe/H]=+0.5 dex, confirming our earlier conclusions that the Galactic center hosts no stars with extreme chemical composition.
\end{abstract}

\keywords{stars: abundances --- nuclear star cluster --- silicon --- late-type --- Galaxy: center}

\section{Introduction}

\myedit{Nuclear star clusters are ubiquitous in galaxies: They are present in $\gtrsim 70\%$ of early and late type galaxies with stellar masses above $10^{8}-10^{10}$\,M$_{\odot}$. at greater stellar masses, their frequency decreases in early type galaxies, but stays high in late types. Nuclear star clusters have been found to coexist with supermassive black holes (SMBH), especially at the low-mass range for SMBH. However, there are numerous cases of nuclear star clusters, where a central black hole could not be detected, i.e. has a tight upper mass limit. The masses of nuclear star clusters display scaling relationships with the masses of their host galaxies. This supports the idea that the nuclei of galaxies may have co-evolved with their hosts.  Reviews by \citet{neumayer:20:arxiv, boker:10, neumayer:2017, seth:19} consider nuclear star clusters, their formation history, and their role in the evolution of galaxies.}

\myedit{The nuclear star cluster at the center of the Milky Way (NSC) has a half-light radius of about 5\,pc (corresponding to about 2 arcmin at the distance of the GC), a mass of $2.5\pm0.4 \times10^{7}$\,M$_{\odot}$, is flattened along the Galactic north-south direction and rotates in parallel to the Galactic Disc. The Milky Way's central black hole, Sagittarius\,A* (Sgr\,A*) is located at the precise center of the NSC. The NSC appears to be very similar to extragalactic NSCs. It is currently the only one where we can observationally resolve individual stars and it therefore plays a key role as a template for its extragalactic counterparts \citep{fritz16,schodel:14,neumayer:20:arxiv}.}

\myedit{The NSC is not well isolated, but lies embedded into the so-called Nuclear Stellar Disk (NSD), a dense stellar region  that has a radius of 100-200\,pc and a scale height of about 45\,pc \citep{launhardt02}. In spite of massive young clusters and other signs of ongoing massive star formation, $80-90\%$ of the stars in the NSD appear to have formed more than 8\,Gyr ago \citep{noguereslara18,nogueraslara:19}.}

\myedit{Due to the co-penetration and overlap of different stellar components along the ling of sight to the GC (disc, bulge/bar, nuclear disc, nuclear cluster) assigning membership of any given star to the NSC is a complex issue. As members of the NSC have orbits that linger at the apocenter, it is not necessary to observe stars very near the black hole to gain insights into the physical properties of the NSC. But contamination of any target sample by stars that are not in the NSC increases as a function of distance from  from Sgr\,A*.}

\myedit{Two channels have been proposed to account for the presence of NSCs in galaxies  \citep[see, e.g.][]{neumayer:2017,neumayer:20:arxiv}. (1) Inspiral via dynamical friction and subsequent merger of globular cluster like systems into the nucleus  \citep{TR75,CD93,AN12}. (2) {\it In situ} formation of stars, that might result from an extended star formation history, including sharing some history with the formation of the bulge/bar population \citep{loose:82,milosavljevi:04,mapelli:12,mastrobuonobattisti:19}.(3) Finally, a channel worthy of consideration would be the capture of Galactic bulge stars on ergodic orbits; these stars might plausibly originate in nuclear discs or spirals or in bars.}

\myedit{In the case of the NSC we can test these scenarios by studying its stars. Very young massive stars detected in the central parsec of the Galaxy provides unambiguous evidence for current in situ star formation \citep{genzel:10,feldmeierkrause:15}. Using galaxy simulations, \citet{guillard:16} proposed a hybrid formation scenario in which clusters bring some gas with them to form a fraction of NSC stars {\it in situ}. If drawn from the Milky Way globular cluster system -- even that of the bulge -- such stars would be expected to contribute stars of predominantly subsolar metallicity. However, young and more metal rich star clusters formed close to the Galactic center may also  have contributed to its build up  \citep{AS15}.}

The compositions of stars can strongly constrain which of these scenarios is dominant. Early investigations in the Galactic center $\sim$tens of parsecs distance from Sgr\,A*, focused on the brightest star in the region, the M2 supergiant IRS 7 \citep{carr:00,davies:09} and later extended to include other bright targets \citep{ramirez2000,cunha:07}.  These studies commondly find  $-0.2 < \rm [Fe/H] <+0.4$  and an \textalpha-abundance in the range between about 0.0 to 0.5\,dex, both in reference to Solar values by \citet{solar:sme}. Analyzing fainter red giants, most likely located more distant in the NSD, \citet{ryde_schultheis:15,ryde:16} find similar results, but with lower alphas.

\citet{do:15} used medium resolution, Adaptive Optics-fed spectroscopy to probe the central parsec, and found some metal poor star candidates and an extremely metal rich abundance distribution reaching up to ten times Solar in [Fe/H].  Metal poor giants were confirmed; the high resolution abundance analysis of a metal poor, alpha-enhanced M giant 58 pc from SgrA* was conducted by \citet{ryde:16:nsc}. \citet{rich:17} reported the first high resolution abundance analysis for members of the NSC, finding a wide abundance distribution with mean [Fe/H]=$-0.16$\,dex extending up to +0.6\,dex. \citet{Feldmeier-Krause2017} used medium resolution KMOS spectroscopy to derive metallicities for 700 stars in the central 4\,pc$^2$.  They find a significantly more metal rich distribution (median 0.29\,dex; extended to +1\,dex). High resolution spectroscopy of 2 stars in the central parsec resulted in a claim by \citet{do:18} for extreme abundances of scandium, vanadium, and yttrium in the NSC, along with iron abundances too high to measure with current atmosphere models. While our work does not have access to the central parsec, we did confirm that the {\it lines} of scandium, vanadium, and yttrium are strong in cool NSC giants \citep{thorsbro:18}. However, from our local disk control sample, we find these lines are also strong in cool local disk giants. Thus, the strong lines are a property of cool stars, and we attribute this strength not to an abundance increase but rather to hyperfine splitting (that delays saturation) and temperature sensitive non-LTE strengthening of the line \citep[for more details, see][]{thorsbro:18}. \myedit{\citet{najarro2009} investigated two luminous blue variables in the Quintuplet cluster and find solar metallicity with enhanced \textalpha-abundances.}

\myedit{Investigations into variable stars are also relevant for the study of the Galactic center as RR Lyrae stars can be tracers of very old stellar populations \citep{minniti:16,dong17} and classical Cepheids can be tracers of young stellar populations \citet{matsunaga:15,kovtyukh:19}. At present, a handful of RR ab variables are found near the NSC from the \citet{dong17} study but their actual membership in the NSC remains a matter of debate.}


In general \textalpha-element abundance trends have not been investigated, which would help constraining different models for the formation and evolution of the inner Galaxy \citep[e.g.][]{matteucci:12,dimatteo:16}.
Here we report a new, interesting abundance trend in \sife, from the observations of 20 M giants in the NSC and vicinity and 7 M giants observed in the Milky Way Disk (MWD). This is a continuation of the work started by \citet{ryde:16:nsc,rich:17,thorsbro:18}.

\begin{deluxetable*}{l c r r r c r c c c }
\tablecaption{Compiled data for the observed stars discussed in this paper. We assume solar abundances of A(Fe)~=~7.45 and A(Si)~=~7.51 \citep{solar:sme}. General uncertainties are listed for each of the stellar parameters and the silicon abundances. The locations are Milky Way Disk (MWD), Nuclear Star Cluster (NSC), and  Stellar Disk (NSD). \label{tab:starsummary}}
\tablewidth{0pt}
\tablehead{
\colhead{Star} & \colhead{K$_\text{s}$} & \colhead{RA} &  \colhead{dec} & \colhead{\teff} & \colhead{\logg} &  \colhead{\feh} & \colhead{$\xi_\mathrm{micro}$} & \colhead{\sife} & \colhead{Location} \\
\colhead{\emph{uncertainties:}} & & & & \colhead{$\pm 150$} & \colhead{$\pm 0.3$} & \colhead{$\pm 0.15$} & \colhead{$\pm 0.15$} & \colhead{$\pm 0.15$} \\   & & \colhead{[h:m:s]} & \colhead{[d:m:s]} & \colhead{[K]} & & \colhead{(dex)} & \colhead{[km\,s$^{-1}$]} & \colhead{(dex)}
} 
\startdata
2M17584888--2351011   & \wo 6.49 & 17:58:48.89 &  $-$23:51:01.17 & 3652 & 1.30 & \wm$0.55$ & 2.0 & \wm$0.06$ & MWD \\
2M18103303--1626220   & \wo 6.51 & 18:10:33.04 &  $-$16:26:22.06 & 3436 & 0.61 & \wm$0.06$ & 2.3 & \wm$0.02$ & MWD \\
2M18191551--1726223   & \wo 6.56 & 18:19:15.51 &  $-$17:26:22.35 & 3596 & 0.84 &   $-0.02$ & 2.2 & \wm$0.00$ & MWD \\
2M18550791+4754062    & \wo 7.63 & 18:55:07.92 & \wm 47:54:06.22 & 3915 & 1.24 &   $-0.20$ & 2.0 & \wm$0.07$ & MWD \\ 
2M19122965+2753128    & \wo 6.60 & 19:12:29.66 & \wm 27:53:12.83 & 3263 & 0.48 & \wm$0.29$ & 2.4 & \wm$0.12$ & MWD \\               
2M19411424+4601483    & \wo 7.69 & 19:41:14.25 & \wm 46:01:48.14 & 3935 & 1.13 &   $-0.41$ & 2.0 & \wm$0.21$ & MWD \\ 
2M21533239+5804499    & \wo 6.58 & 21:53:32.40 & \wm 58:04:49.94 & 3708 & 1.28 & \wm$0.35$ & 2.0 &   $-0.01$ & MWD \\
GC 25\,\tablenotemark{a} & 10.44 & 17:45:40.93 &  $-$29:00:24.39 & 3383 & 0.38 &   $-0.08$ & 2.5 & \wm$0.08$ & NSC \\
GC 6227               &    11.82 & 17:45:38.86 &  $-$29:01:08.71 & 3681 & 0.92 &   $-0.27$ & 2.1 & \wm$0.11$ & \,~\,\,NSC\,\tablenotemark{b} \\
GC 7104               &    10.74 & 17:45:38.94 &  $-$29:00:58.44 & 3657 & 1.09 & \wm$0.30$ & 2.0 & \wm$0.20$ & NSD \\
GC 8623               &    10.53 & 17:45:41.83 &  $-$29:00:46.15 & 3433 & 0.38 &   $-0.12$ & 2.5 & \wm$0.07$ & NSC \\
GC 8631               &    11.56 & 17:45:43.02 &  $-$29:00:46.02 & 3899 & 1.41 & \wm$0.27$ & 1.9 & \wm$0.23$ & NSC \\
GC 10195              &    10.67 & 17:45:43.10 &  $-$29:00:25.45 & 3414 & 0.55 &   $-0.01$ & 2.4 & \wm$0.03$ & \,~\,\,NSC\,\tablenotemark{b} \\
GC 10812              &    10.25 & 17:45:37.23 &  $-$29:00:16.62 & 3895 & 0.60 &   $-1.02$ & 2.3 & \wm$0.41$ & NSD \\
GC 11025              &    10.41 & 17:45:37.13 &  $-$29:00:14.39 & 3359 & 0.64 & \wm$0.25$ & 2.3 & \wm$0.25$ & NSC \\
GC 11473              &    11.74 & 17:45:42.64 &  $-$29:00:10.23 & 3520 & 1.06 & \wm$0.38$ & 2.0 & \wm$0.26$ & NSD \\
GC 11532              &    10.90 & 17:45:42.90 &  $-$29:00:09.60 & 3450 & 0.48 &   $-0.23$ & 2.4 & \wm$0.09$ & NSC \\
GC 13282              &    10.99 & 17:45:39.49 &  $-$28:59:58.74 & 3543 & 0.77 &   $-0.04$ & 2.2 &   $-0.02$ & NSC \\
GC 13727              &    10.82 & 17:45:39.59 &  $-$28:59:56.21 & 3382 & 0.31 &   $-0.28$ & 2.5 & \wm$0.15$ & NSC \\
GC 13882              &    10.56 & 17:45:38.70 &  $-$28:59:55.15 & 3344 & 0.35 &   $-0.17$ & 2.5 & \wm$0.01$ & NSC \\
GC 14024              &    10.99 & 17:45:42.03 &  $-$28:59:54.97 & 3450 & 0.96 & \wm$0.49$ & 2.1 & \wm$0.15$ & NSC \\
GC 14724              &    10.56 & 17:45:37.31 &  $-$28:59:49.43 & 3482 & 0.45 &   $-0.32$ & 2.4 & \wm$0.14$ & NSC \\
GC 15540              &    10.49 & 17:45:41.93 &  $-$28:59:23.39 & 3496 & 0.61 & \wm$0.01$ & 2.3 & \wm$0.04$ & NSC \\
GC 16763              &    10.66 & 17:45:37.29 &  $-$29:01:12.67 & 3510 & 0.58 &   $-0.15$ & 2.3 &   $-0.03$ & NSC \\
GC 16867              &    11.77 & 17:45:36.02 &  $-$29:00:09.20 & 3628 & 0.76 &   $-0.18$ & 2.2 & \wm$0.02$ & NSC \\
GC 16887              &    11.63 & 17:45:44.04 &  $-$28:59:27.66 & 3308 & 0.77 & \wm$0.41$ & 2.2 & \wm$0.18$ & NSC \\
GC 16895              &    10.75 & 17:45:35.64 &  $-$29:00:47.00 & 3438 & 0.50 &   $-0.13$ & 2.4 & \wm$0.07$ & NSC \\
\enddata
\tablenotetext{a}{GC 25 appears in other works and is listed as \emph{Id=44} in \citet{Feldmeier-Krause2017} and \emph{Name=NE1-1 001} in \citet{do:15,stostad:17}. Note that the star listed as GC 25 in \citet{nandakumar:18} is not the same star in spite of having the same name.}
\tablenotetext{b}{GC 6227 and GC 10195 are possible background stars depending on whether they are intrinsically reddened.}
\end{deluxetable*}

\section{Observations}

27 M giants have been observed at medium/high spectral resolution in the K-band using the NIRSPEC \citep{nirspec_mclean,mclean} facility at Keck~II, and are summarized in Table~\ref{tab:starsummary}. We used the $0.432'' \times 12''$ slit and the NIRSPEC-7 filter, giving the resolving power of $R\sim23,000$, which is a minimum for determining accurate abundances from spectra of such cool stars. Five spectral orders are recorded, covering the wavelength range of 21,000-24,000\,\AA. However, the wavelength coverage is not complete; there are gaps between the orders. 

\begin{figure*}
\centering
\includegraphics[trim={0cm 0cm 0cm 0cm},clip,angle=0,width=1.00\hsize]{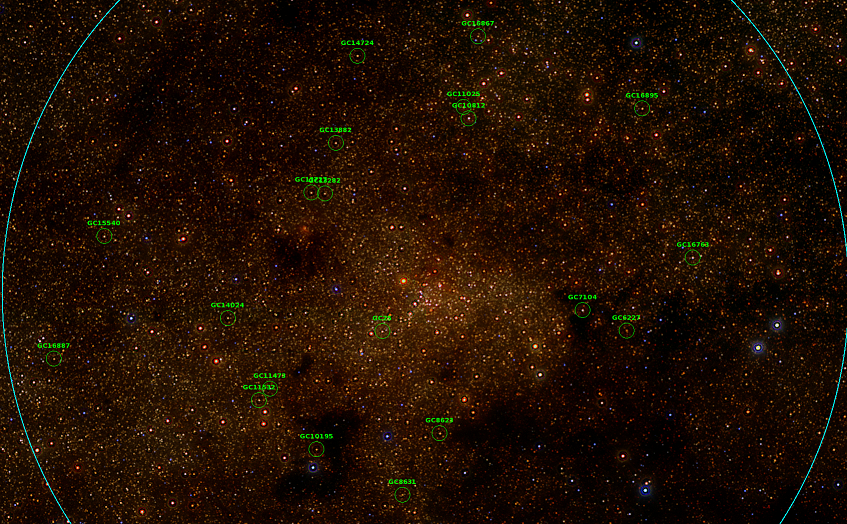}
\caption{A JHKs RGB image from the GALACTICNUCLEUS survey \citep{noguereslara18,GALACTICNUCLEUS:19} marking the observed stars in the region inside 3.5\,pc projected distance from Sgr\,A* illustrated with the blue circle.} \label{fig:findingchart}
\end{figure*}

\begin{figure}
\centering
\includegraphics[trim={0cm 0cm 0cm 0cm},clip,angle=0,width=1.00\hsize]{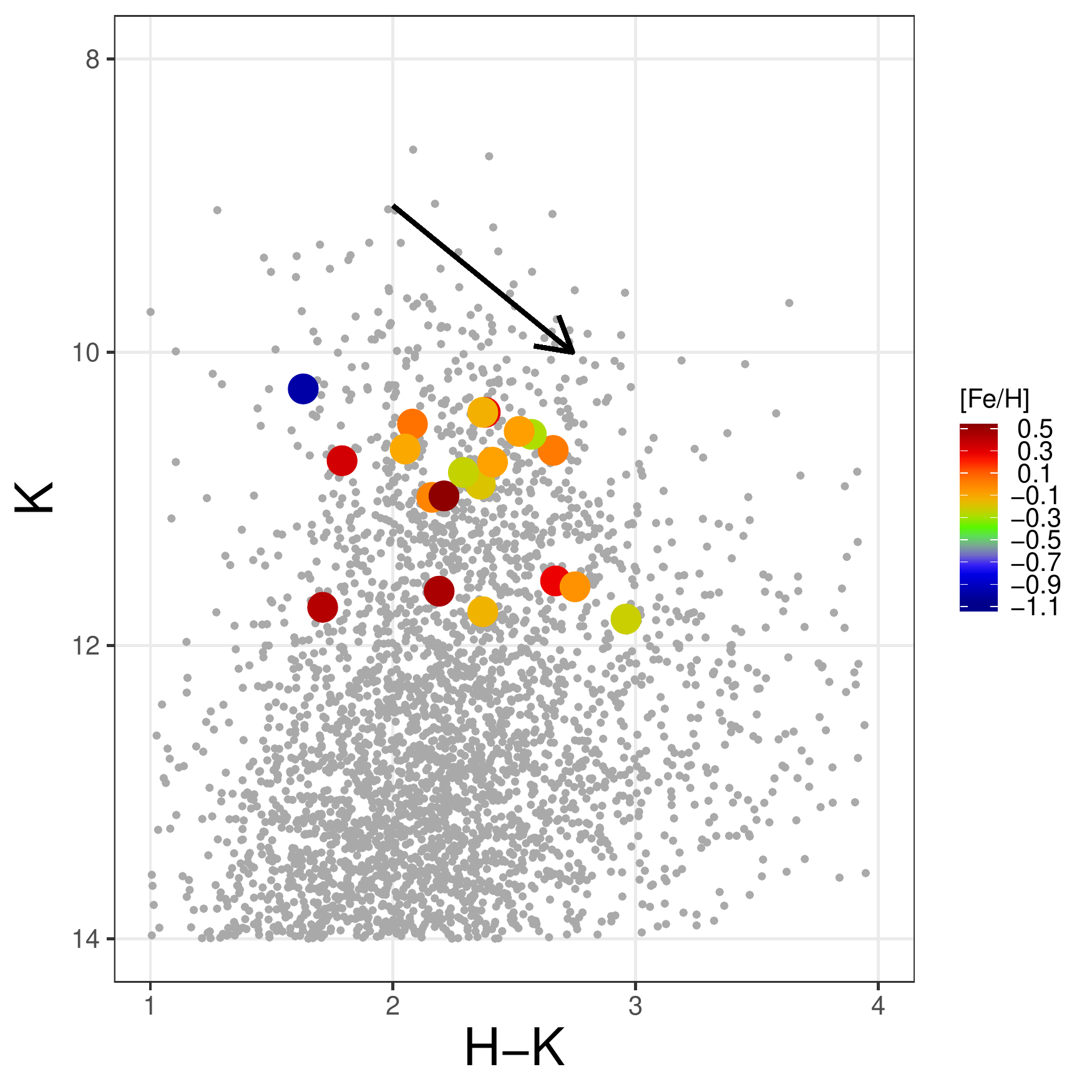}
\caption{2MASS K vs H-K diagram in the NSC superimposed by our sample of 20 M giants colored by metallicity (see Sect.~\ref{metallicity}).} \label{fig:CMD}
\end{figure}

Seven of these observed stars are located in the MWD with similar temperature range to the observed Galactic center stars, thus enabling them to be used as a control sample having similar systematic uncertainties. The MWD stars are presented in \citet{thorsbro:18}, where they are important for the discussion on the strong, K-band Sc\,{\sc I} lines detected in Galactic center stars. The other 20 stars are located in the Galactic center, of which 17 are observed under the programs U87NS (April 2015, PI: Rich) and U103NS (April 2016, PI: Rich) are presented in \citet{ryde:16,rich:17}. Figure~\ref{fig:findingchart} shows a JHKs RGB image from the GALACTICNUCLEUS survey \citep{GALACTICNUCLEUS:19} of the observed stars. Figure~\ref{fig:CMD} shows a color-magnitude diagram (CMD) with our Galactic center M giants, color-coded with the derived metallicities (see Sect.~\ref{metallicity}), superimposed on the 2MASS sample of the Galactic Center field. Our target stars have been selected over the full width of the CMD in order to avoid any selection bias. Note that the star listed as GC 7688 in \citet{rich:17} is not included as it has been identified as a foreground bulge star. Three stars have been added to the sample, all of them observed on 2017 July 28-29, under program U103NS (PI: Rich). These are selected in the same way as the 17 original stars from low resolution SINFONI spectra \citep[see][]{rich:17}. Similarly, these are observed with an ABBA scheme with a nod throw of $6\arcsec$ along the slit, in order to achieve proper background and dark subtraction. The three added stars have each been exposed for 2000\,s. These stars are also reduced with the NIRSPEC software {\tt redspec} \citep{nirspec_reduction}, and thereafter with IRAF \citep{IRAF} to normalize the continuum, eliminate cosmic ray hits, and correct for telluric lines. The latter has been done with a high signal-to-noise spectrum of the rapidly rotating O6.5V star HIP89584. For more details about the data reduction process, we refer to \citet{rich:17}.

\section{Analysis}

Our goal is to determine the \afe-ratios as a function of metallicity for stars in the NSC and its vicinity. However, the spectra of the Galactic center stars show a surprising amount of features due to the presence of diatomic molecules, in particular the CN molecule. To perform a good abundance analysis it is important to make sure all spectral features are identified so a line list containing unblended lines can be constructed for good abundance analysis, see Section~\ref{linelist}.

We analyze the M giant spectra using tailored synthetic spectra, calculating the radiative transfer through spherical model atmospheres. In order to derive as accurate abundances as possible, the fundamental input parameters for these stellar atmospheres are needed to be well determined. These parameters are the effective temperature, \teff, surface gravity, \logg, metallicity, \feh, and the microturbulence, \micro, of the stars. These are discussed in Sections \ref{temp}-\ref{metallicity}.

In an earlier study, we analyzed a giant with [Fe/H]$\sim -1$ found in the vicinity of the NSC \citep{ryde:16}, measuring abundances for Mg, Si, Ca, and Ti. The remaining NSC giants are cooler and more metal rich; molecules and blending severely limits the usable lines so that at our resolution so we are able only to measure the \textalpha-element Si (see  Section~\ref{silicon}). 

From three KECK runs, during the period 2015-2017, we have gathered spectra of a sample of 20 M giants that are all shown to be confined to the NSC and near vicinity (see Section~\ref{orbit}).


\subsection{Line List}\label{linelist}

In order to synthesize the spectra, an accurate line list is required. In the list provided by \citet{thorsbro:17} wavelengths and line-strengths (astrophysical $\log gf$-values) are updated using the solar center intensity atlas \citep{solar_IR_atlas}. We have also made use of recent laboratory measurements of atomic line strengths of Mg\,{\sc I}, and Sc\,{\sc I} \citep{pehlivan:mg,pehlivan:sc}, as well as of Si\,{\sc I} (Pehlivan Rhodin et al., 2020, submitted) in the K band. Of approximately 700 identified, interesting spectral lines for cool stars, about 570 lines have been assigned new values.

As molecular lines have a crucial impact in terms of blending, we include molecular lines in the line list. For CN we include the list from \citet{sneden:16}, which is the most dominant molecule apart from CO, whose lines dominate  in the  2.3\,\textmu m bandhead  region. The CO line data are from \citet{goor}. At the shorter wavelengths of our spectra region SiO, H$_2$O, and OH are important. Line lists for these molecules are included into our line list from respectively \citet{langhoff:07,Barber06,brooke:16}.

\begin{figure}
\centering
\includegraphics[trim={4.1cm 11.4cm 2.0cm 12cm},clip,angle=0,width=1.00\hsize]{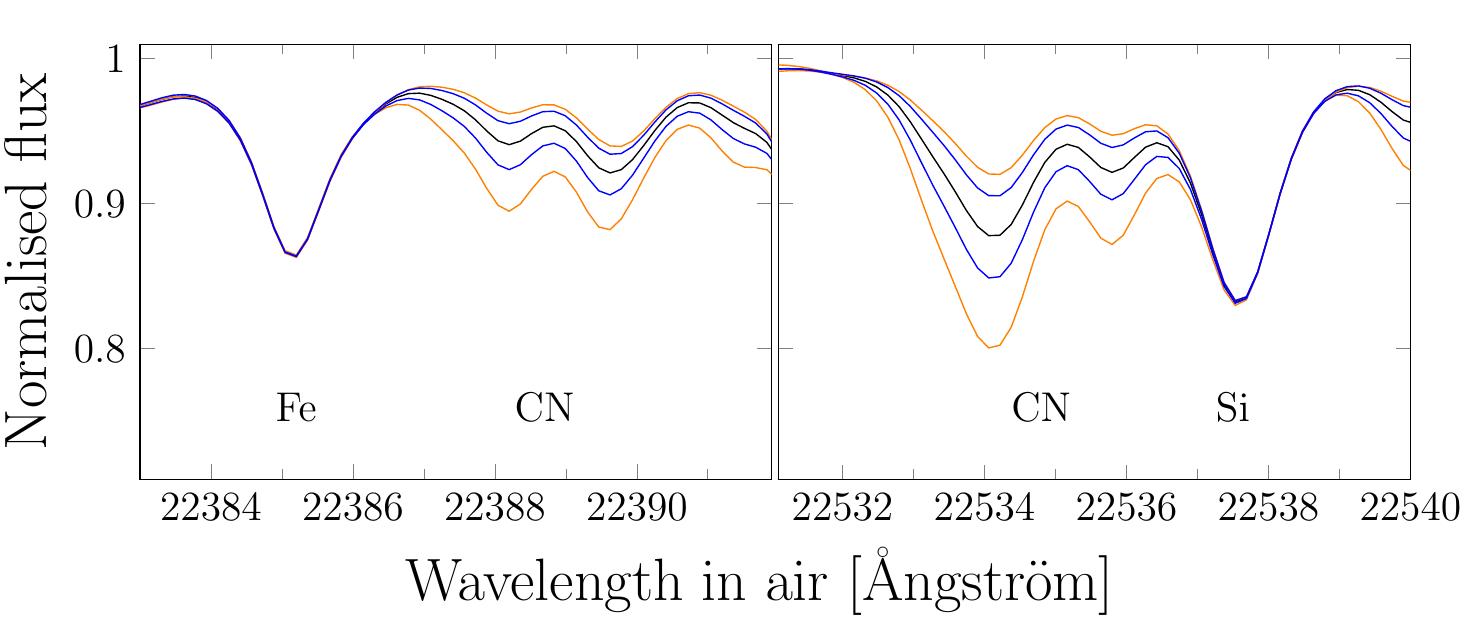} 
\caption{Synthetic model spectra for a star with solar metallicity at 3600 K effective temperature. The black line shows typical M giant abundances of carbon and nitrogen, the blue lines show the effect of varying the nitrogen abundance by $\pm0.2$\,dex, and the orange lines show a variation in the carbon abundance of $\pm0.2$\,dex. The Fe\,{\sc I} line shown in the left plot and the Si\,{\sc I} line in the right plot are not blended with CN molecules. For the stars we are investigating, CN blends are very common in the near infrared, so finding unblended lines is crucial. \label{fig:sensi}}
\end{figure}

To identify clean spectral lines suitable for measurement, we examine synthetic model spectra of cool M~giants to identify iron lines that are not blended with molecular features. The CN line list used \citep{sneden:16} has been shown to reproduce the observed CN spectrum very well, and can be used to identify blends. Unknown blends might still always be present and special care has been taken when such possible mismatches are identified. Examples of lines that are not blended are illustrated in Figure ~\ref{fig:sensi}, where an Fe\,{\sc I} line and a Si\,{\sc I} line are shown. 

For the determination of Fe and Si abundances the lines shown in Table~\ref{tab:lines} have been identified as good lines with little or no blending.
\begin{deluxetable}{L C C C }
\tablecaption{Lines used for abundance determinations. \label{tab:lines}}
\tablewidth{1.0\columnwidth}
\tablehead{
\colhead{Wavelength in air}  & \colhead{$E_\mathrm{exc}$} & \colhead{$\log gf$}   \\
 \colhead{[\AA] } & \colhead{[eV]} & \colhead{} }
\startdata
{\rm Fe\,{\sc I}} & & \\
21178.211 & 3.0173 & -4.201 \\
21238.509 & 4.9557 & -1.281 \\
21284.348 & 3.0714 & -4.414 \\
21851.422 & 3.6417 & -3.506 \\
22380.835 & 5.0331 & -0.409 \\
22385.143 & 5.3205 & -1.536 \\
22392.915 & 5.0996 & -1.207 \\
22419.976 & 6.2182 & -0.226 \\
\hline
{\rm Si\,{\sc I}} & & \\
21354.249 & 6.2228 & {\color{white}-}0.138 \\
21368.693 & 7.3288 & -1.905 \\
21874.199 & 6.7206 & -0.731 \\
21879.345 & 6.7206 & {\color{white}-}0.384\\
21924.981 & 7.0645 & -1.269 \\
22537.593 & 6.6161 & -0.216 \\
\enddata
\end{deluxetable}

\subsection[Effective Temperature]{Effective Temperature, \teff}\label{temp}

The effective temperatures of the stars in the MWD are obtained following  \citet{thorsbro:18}, where the temperatures from the APOGEE Data Release 14 \citep{majewski:17,abolfathi:18,blanton:17} were examined and then adopted.


To determine the temperatures of the stars in the Galactic center we rely on the method of
\citet{schultheis:16} as employed in \citet{rich:17}, which is based on using low-resolution SINFONI \citep{Eisenhauer_03} spectra and measures an index of the CO band head at $2.3$\,\textmu m. \citet{rich:17} determined the effective temperatures for 17 of the 20 Galactic center stars in our sample. The same method is used for determining the temperature for the added stars, namely GC 8623 and GC 13727. We do not have SINFONI spectra for the GC 25 star and thus it remains undetermined using the CO band head method. The CO band head temperatures are shown in Table~\ref{tab:teff} in the ``CO band'' column.
\begin{deluxetable}{l C C C}
\tablecaption{Analysis of the effective temperature of stars in the Galactic center. \label{tab:teff}}
\tablewidth{1.0\columnwidth}
\tablehead{
\colhead{Star}  & \colhead{CO band} & \colhead{Scandium} & \colhead{Mean} \\
 & \colhead{$T_{\mathrm{eff}}$} & \colhead{$T_{\mathrm{eff}}$} & \colhead{$T_{\mathrm{eff}}$} \\
 & \colhead{[K]} & \colhead{[K]} & \colhead{[K]}}
\startdata
GC 25     & ... & 3383 & 3383 \\ %
GC 6227   & 3800 & 3561 & 3681 \\ 
GC 7104   & 3750 & 3563 & 3657 \\ 
GC 8623   & 3415 & 3451 & 3433 \\ %
GC 8631   & 3900 & 3897 & 3899 \\ %
GC 10195  & 3500 & 3327 & 3414 \\ %
GC 10812  & 3800 & 3990 & 3895 \\ %
GC 11025  & 3400 & 3318 & 3359 \\ %
GC 11473  & 3550 & 3489 & 3520 \\ 
GC 11532  & 3500 & 3399 & 3450 \\ %
GC 13282  & 3700 & 3386 & 3543 \\ 
GC 13727  & 3372 & 3391 & 3382 \\ 
GC 13882  & 3300 & 3387 & 3344 \\ %
GC 14024  & 3650 & 3249 & 3450 \\ 
GC 14724  & 3600 & 3364 & 3482 \\ 
GC 15540  & 3600 & 3391 & 3496 \\ 
GC 16763  & 3600 & 3419 & 3510 \\ 
GC 16867  & 3650 & 3606 & 3628 \\ %
GC 16887  & 3500 & 3116 & 3308 \\ 
GC 16895  & 3450 & 3425 & 3438 \\ %
\enddata
\end{deluxetable}

To decrease the uncertainty in the determined temperatures for the stars in the Galactic center and to increase the correlation between the temperatures of the stars in the MWD stars and the stars in the Galactic center, we use the fact that there are strong Sc\,{\sc I} lines in the K-band spectrum that are very temperature sensitive in the 3000-4000 K range \citep[e.g.][]{thorsbro:18}. The effects of different scandium abundances, surface gravities or metallicities are a second order effect. Most of our observed stars are within this temperature range and the strengths of these lines should therefore be able to be used as a temperature indicator. We measure the equivalent widths for a set of scandium lines in the MWD stars for which the effective temperatures are known. We create four linear relations between the effective temperature and the equivalent widths of the four scandium lines at the following wavelengths in air: 21730~Å, 21812~Å, 21842~Å, and 22394~Å. Using the equivalent widths measured of the same scandium lines in the spectra of the Galactic center stars together with these empirical relations, and taking the average of the four results \myedit{(having standard deviations on the order of 50\,K)}, we can thus determine the effective temperature of the Galactic center stars. The effective temperatures determined this way are given in Table~\ref{tab:teff} under the column ``Scandium''.

Comparing the effective temperature values found in Table~\ref{tab:teff}, and with a general uncertainty of 150 to 200\,K, we find that the results from the different methods are in general agreement. We decide to take an average of the values, and keep a conservative estimate of 150\,K uncertainty for the effective temperature values. These temperatures are given in the ``Mean'' column in Table \ref{tab:teff}. For GC~25, we use the temperature determined from the Sc\,{\sc I} lines.

\subsection[Surface Gravity and Microturbulence]{Surface Gravity, \logg, and Microturbulence, \micro}\label{logg}

\citet{rich:17} demonstrated that in spite of the distance to the stars in the \myedit{Galactic center} is known to a good degree, the photometrically determined surface gravities are plagued with large uncertainties due to the patchy extinction. To determine surface gravity, \logg, we instead benefit from the fact that isochrones of old ($>1$ Gyr) M giants are quite insensitive to the age of the star. This reduces isochrones to depend on surface gravity, effective temperature, and metallicity. We thus use a grid of Yonsei-Yale isochrones \citep{demarque:04} to determine \logg\ given an effective temperature and metallicity. The construction and validation of this method is described in more detail in \citet{rich:17}.

The microturbulence, $\xi_\mathrm{micro}$, that takes into account the small scale, non-thermal motions in the stellar atmospheres, is important for saturated lines, influencing their line strengths. We estimate this parameter from an empirical relation with the surface gravity based on a detailed analysis of spectra of five red giant stars ($0.5 < \log g < 2.5$) by \citet{smith:13}, as described in \citet{rich:17}.

The estimated uncertainty of 150\,K for the effective temperatures propagates into an uncertainty of 0.3 for \logg\ and 0.15\,km\,s$^{-1}$ for $\xi_\mathrm{micro}$.

\subsection{Metallicity, \feh}\label{metallicity}

The metallicities of our stars are important not only for identifying the model atmosphere for a certain star but also for the Metallicity Distribution Function (MDF), which is an important diagnostic for the discussion of the origin of the Nuclear Star Cluster, see Sect.~\ref{disc}. 

In order to determine metallicities we synthesize spectra and compare them to the observed spectra. We have chosen to use the spectral synthesis code Spectroscopy Made Easy (SME) \citep{sme,sme_code} which interpolates in a grid of one-dimensional (1D) MARCS atmosphere models \citep{marcs:08}. These are hydrostatic model atmospheres in spherical geometry, computed assuming LTE, chemical equilibrium, homogeneity, and conservation of the total flux (radiative plus convective, the convective flux being computed using a mixing-length recipe). This code has the advantages that it includes a flexible $\chi^2$ minimization tool for finding the solution that fits an observed spectrum the best in a pre-specified spectral window. It also has a powerful continuum normalization routine. In cool star spectra, extra care is needed to normalize the spectrum in the region of a spectral line under consideration. The ubiquitous molecules and unavoidable residuals from the telluric line division necessitates a careful analysis line by line.

We recognize that using 3D atmospheric models would be an improvement over 1D atmospheric models \citep{scott:15:i,scott:15:ii}. However, for this differential study we expect to see the same systematics to be present in both the Galactic center and the MWD stellar samples.

\begin{figure}
  \centering
\includegraphics[trim={0cm 0cm 0cm 0cm},clip,angle=0,width=1.00\hsize]{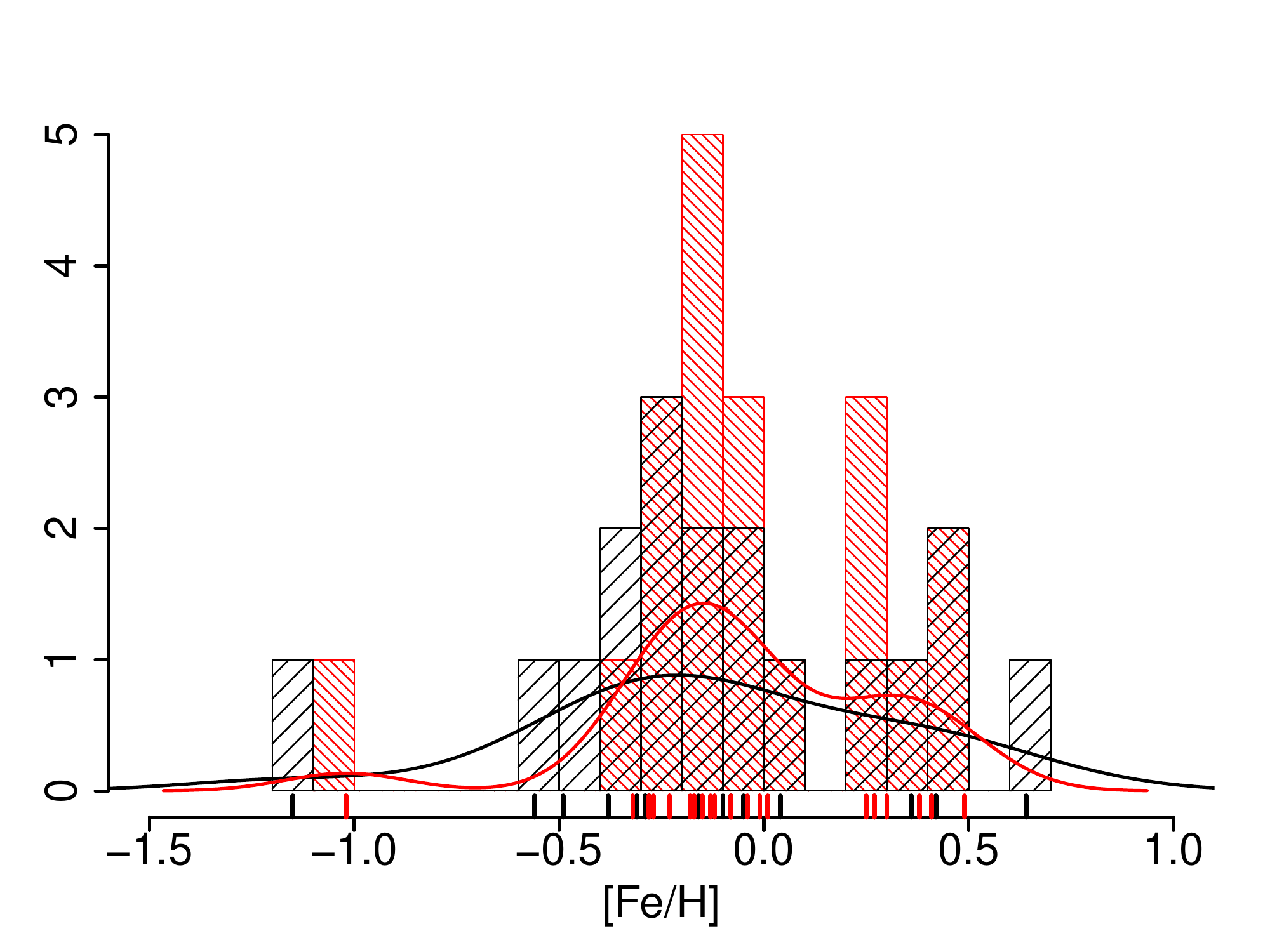} \caption{Determined metallicity distribution function (MDF) in red compared to that of \citet{rich:17} in black. The reanalyzes of the stellar parameters in this work does not significantly change the MDF of the Nuclear Stellar Cluster.}
\label{fig:MDF:rich}
\end{figure}

As a result of our updated temperatures and careful work on the line list, especially the new analysis of lines blended with CN, we reanalyze the [Fe/H] values given in  \citet{rich:17}. The new [Fe/H] values are shown in Table~\ref{tab:starsummary}, with a general estimated uncertainty of 0.15\,dex. A comparison of our new [Fe/H] distribution function (MDF) with that of \cite{rich:17} is shown in Figure ~\ref{fig:MDF:rich}.

Principally due to our new treatment of CN blending, our iron abundances have changed compared to earlier work; for example, the most metal rich star now is [Fe/H]=+0.5, compared to +0.64 in \citet{rich:17}.  We consider the present abundance distribution reported in this work to be the best current measurement of [Fe/H] and its range in the NSC.


\subsection[Alpha-element Abundance Determination]{$\alpha$-element Abundance Determination}\label{silicon}

The \textalpha-element abundances can potentially be measured from several lines from different species. However, considering the line strengths and number of unblended lines, there are only a few elements possible in the available wavelength region at the spectral resolution considered. The available Mg\,{\sc I} lines are heavily blended with CN molecular lines and the strengths of the Ti\,{\sc I} and Ca\,{\sc I} lines increase conspicuously with cooler effective temperature, broadly similar to the Sc\,{\sc I} lines discussed in detail by \citet{thorsbro:18}. This behavior argues strongly for non-LTE effects affecting the line strengths more than the abundances. Since no non-LTE corrections are available for these elements in the temperature range that our stars span, these lines cannot yet be used for determining the abundances. Si\,{\sc I}, however, has several suitable lines and most importantly, non-LTE departure coefficient calculations show that departure from LTE is not important for Si\,{\sc I} lines in the stellar parameter space of the stars in our sample. We verified this by examining the departure coefficients in MARCS model giants stars, using the atomic model presented in \cite{amarsi:17}, which utilizes the rather efficient inelastic hydrogen collisions of \citep{belyaev:14}. 

\begin{figure}
\centering
\includegraphics[trim={0cm 0cm 0cm 0cm},clip,angle=0,width=1.00\hsize]{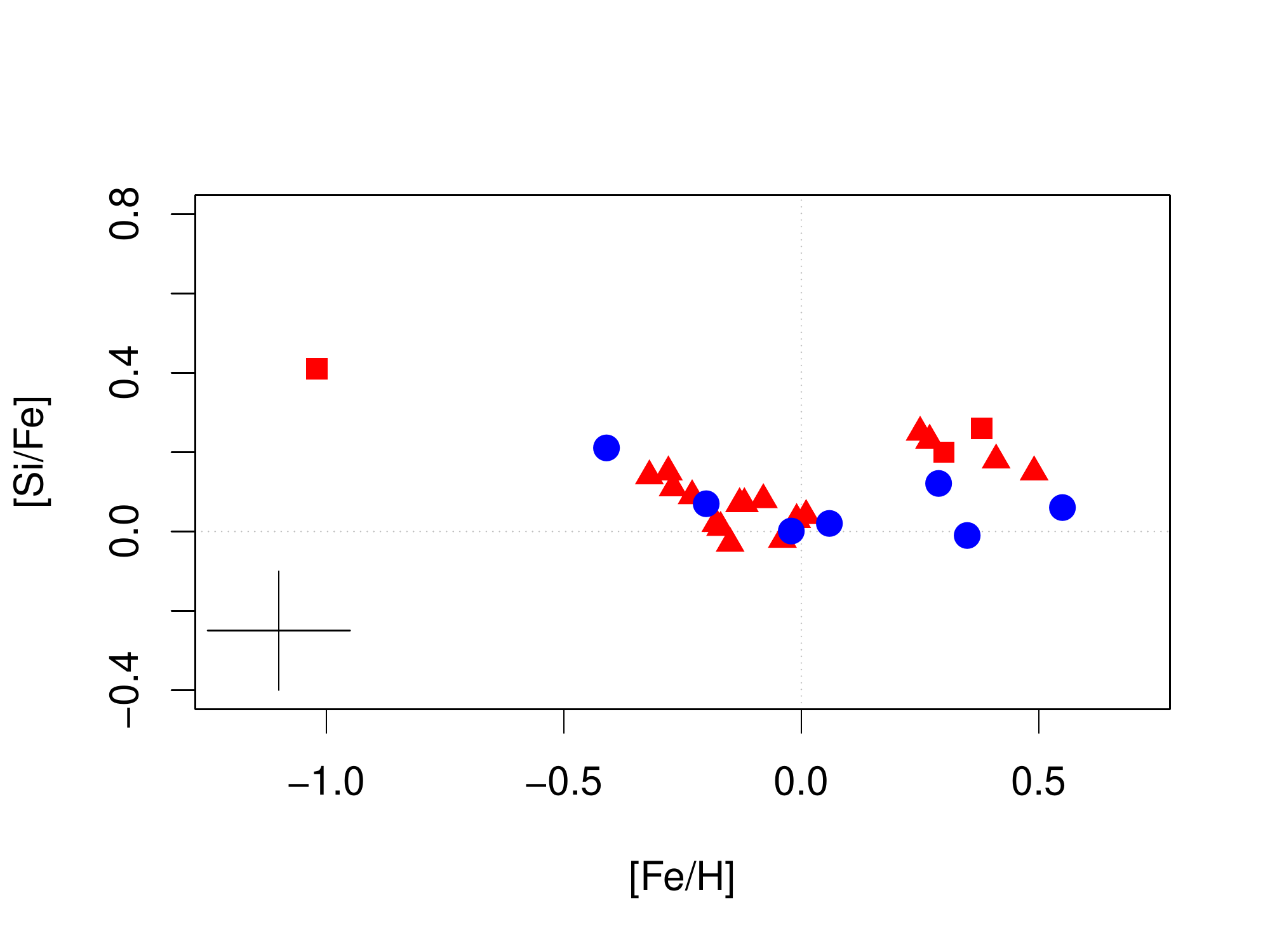} 
\caption{[Si/Fe] vs. [Fe/H]. Red triangles and squares are Nuclear Star Cluster respective Nuclear Stellar Disk stars, blue disks the Milky Way Disk stars.} \label{fig:sife}
\end{figure}

\begin{figure}
\centering
\includegraphics[trim={4.1cm 11.4cm 2.0cm 12cm},clip,angle=0,width=1.00\hsize]{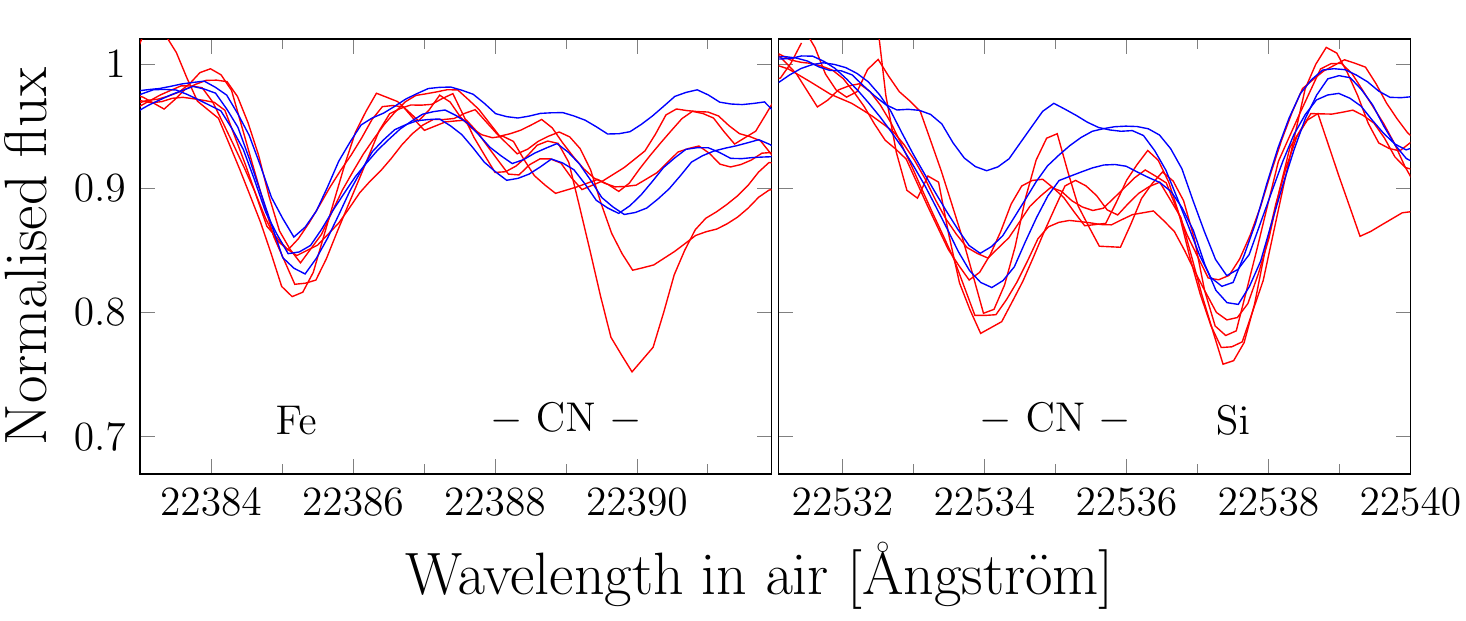} 
\caption{Observed spectra of the subset of stars from our sample with super solar metallicities showing an unblended Fe\,{\sc I} line and an unblended Si\,{\sc I} line. Blue spectra from Milky Way Disk stars, and red spectra for Galactic center stars. The Si\,{\sc I} lines are visibly deeper for the Galactic center stars compared to the Milky Way Disk stars, while the Fe\,{\sc I} lines show negligible variation. \label{fig:obssilicon}}
\end{figure}

The \textalpha-element abundance is therefore determined from the unblended Si\,{\sc I} lines available (Table \ref{tab:lines}) with a general estimated uncertainty of 0.15\,dex. Typical spectra are shown in Figure \ref{fig:obssilicon}.  The silicon abundances are shown in Figure~\ref{fig:sife} with the Galactic center stars marked with red triangles and squares, the latter for those not identified to be in the NSC (see Section~\ref{orbit}) and the MWD stars with blue disks.

We now examine more closely those metal rich stars with Si enhancement. Figure~\ref{fig:obssilicon} shows our observed spectra for the metal rich subset giving an extra sanity check on the abundance results. The plot indicates stronger silicon features when comparing the Galactic center stars to the MWD stars, but the Fe\,{\sc I} line does not show the strong difference we see in the Si\,{\sc I} line. This figure shows that, independent of any abundance analysis, we can see that Si appears to be enhanced in the Galactic center for this subset of stars.


\subsection{Stellar dynamics and positions}\label{orbit}

We examine the stellar dynamics of the three stars added to our data set to verify if they belong to the NSC. Secondly, we also check with ``metal rich'' stars, [Fe/H]$>$0.1, differ in their position or dynamics. 

\begin{figure}
\centering
\includegraphics[trim={0cm 0cm 0cm 0cm},clip,angle=0,width=1.0\hsize]{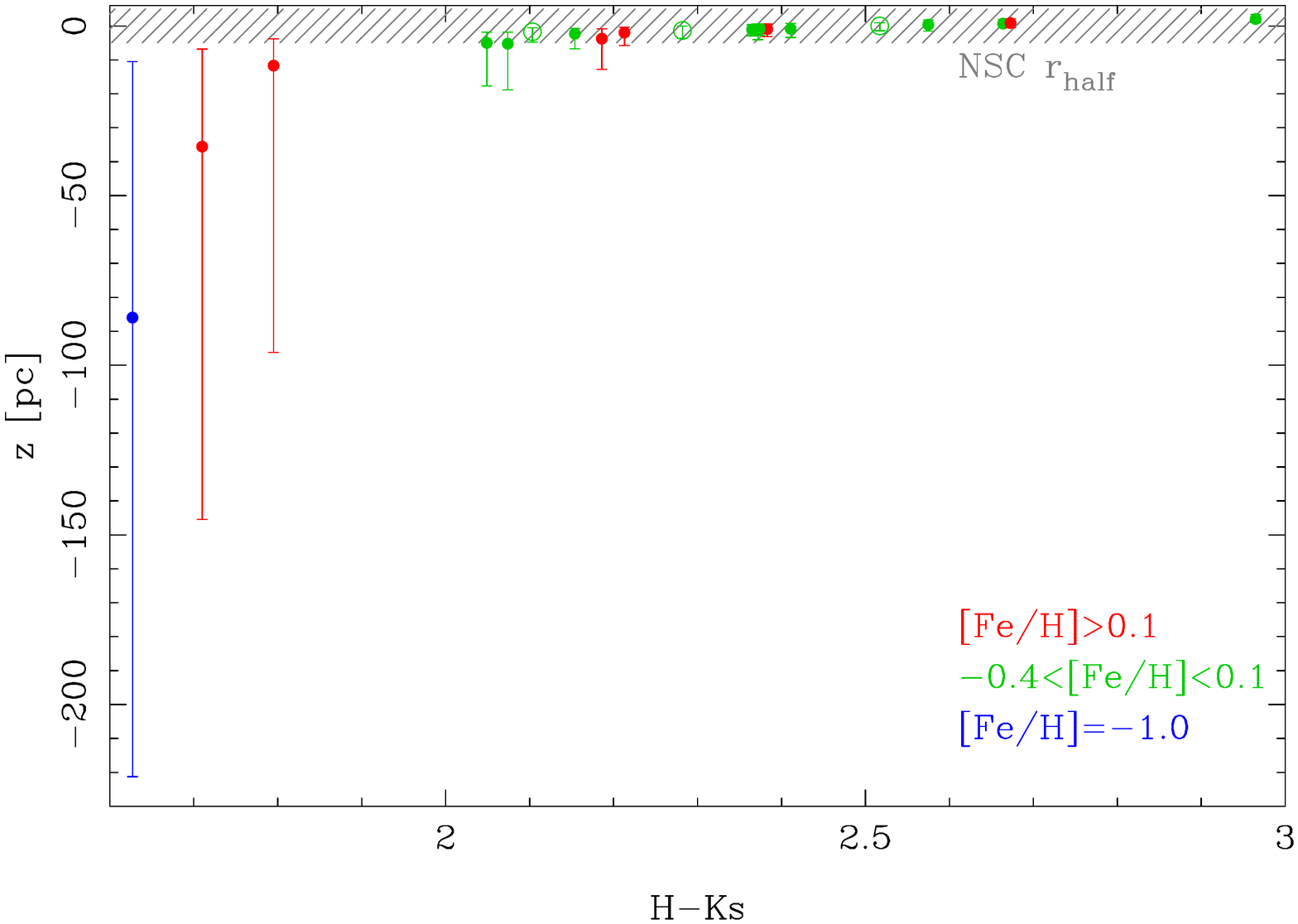} 
\\[-1.0cm]
\includegraphics[trim={0cm 0cm 0cm 0cm},clip,angle=0,width=1.0\hsize]{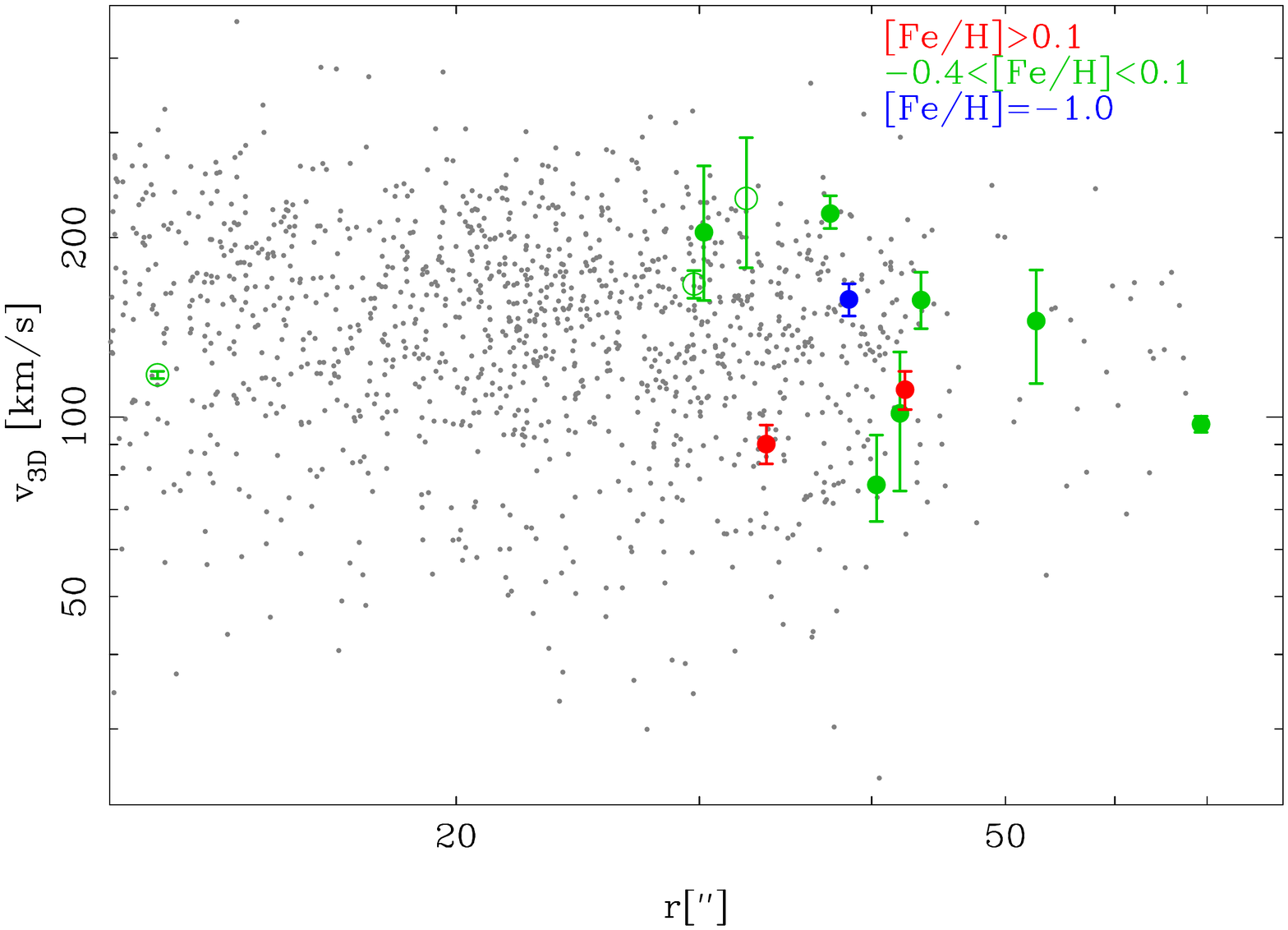} 
\\[-1.0cm]
\includegraphics[trim={0cm 0cm 0cm 0cm},clip,angle=0,width=1.0\hsize]{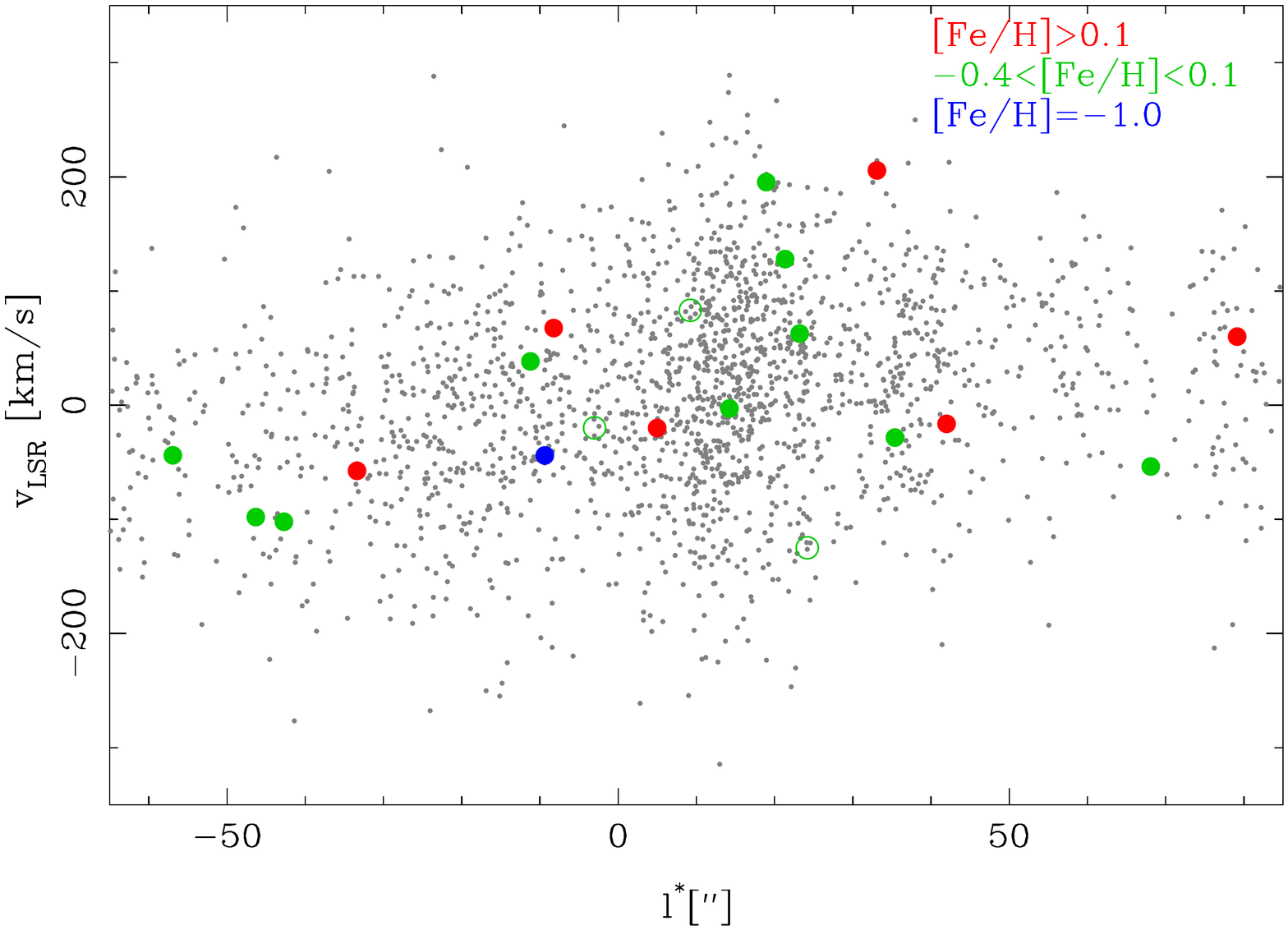} 
\caption{Comparison of sample stars in dynamic and position properties, in all plots the three new stars are plotted as triangles and the old as solid circle. The color indicates the metallicity group. \textit{Top:} The line-of-sight position of all stars compared to the NSC. The NSD covers the full displayed range. \textit{Middle:} Total velocity; not all stars are shown here, because some are missing proper motions. \textit{Bottom:} LSR corrected line-of-sight velocity versus Galactic longitude relative to Sagr~A*. Only stars outside the central 10 arcseconds are shown, since there the black hole increases velocities.} \label{fig:dynamics}
\end{figure}

As in \citet{ryde:16:nsc} and \citet{rich:17} we use the H-K color which is nearly only caused by extinction  for testing whether the stars are currently in the NSC in the line-of-sight direction. We find that the $v_\text{rad}^\text{LSR}$ for the added stars GC~25, GC~8623 and GC~13727 are respectively $83$, $-20$ and $-125$ km\,s$^{-1}$ from the Keck spectra. These velocities are consistent with the lower precision values in \citet{fritz16}. For the proper motions we use \citet{fritz16} as previously. The new stars are all likely members of the NSC, see Figure~\ref{fig:dynamics}. 

As visible in Figure~\ref{fig:dynamics} we do not find differences between the three added stars or the different metallicity groups. All stars seem to belong to the same hot system with a weak rotation aligned with the Galaxy. The sample, especially of stars with proper motions, is still too small to detect minor differences. We note that three stars, GC 10812, GC 7104 and GC 11473, have a preferred location in the NSD, of which the latter two are in the metal rich set. Two of the stars GC, 6227 and GC 10195, could possible be background stars, depending of if they are intrinsically reddened. While effects like motion can lead to exchange between the NSC and the NSD, it is still prudent to account for the possibility that some of these stars may reside in the NSD when explaining the enrichment pattern. However, there is still a need for a Galaxy model which includes all four central components (Sgr\,A*, the NSC, the NSD, and the bar/bulge) to better constrain the location of the stars. Mixing of stars is also a concern to take note of, at the projected distance of 2\,pc from Sgr\,A* there may be about 1 NSD star for every 3-4 NSC stars \citep{gallegocano:20}.

In the line of sight it seems that there could be a larger metallicity range in the NSD than in the NSC, since of the three stars there, none is a normal metallicity stars. Given the low number of stars, it could be chance also, since \citet{do:15} and \citet{Feldmeier-Krause2017} find metal poor stars  which are at least projected NSC members.

\section{Results and Discussion}\label{disc}

A new analysis of iron lines including the most comprehensive currently available CN line list confirms that our sample's most metal rich star has [Fe/H]$\,\sim+0.5$. Our work continues the trend of recent work \citep{ryde_schultheis:15,rich:17,nandakumar:18} that finds stars with metallicity higher than Solar, but finds no evidence in the NSC for extreme metallicities. We therefore can not confirm the assertion of \citet{do:15,Feldmeier-Krause2017} that stars in the NSC have [Fe/H] at or very close to  +1\,dex.

\begin{figure}
  \centering
\epsscale{1.00}
\includegraphics[trim={0cm 0.7cm 0cm 1.5cm},clip,angle=0,width=1.00\hsize]{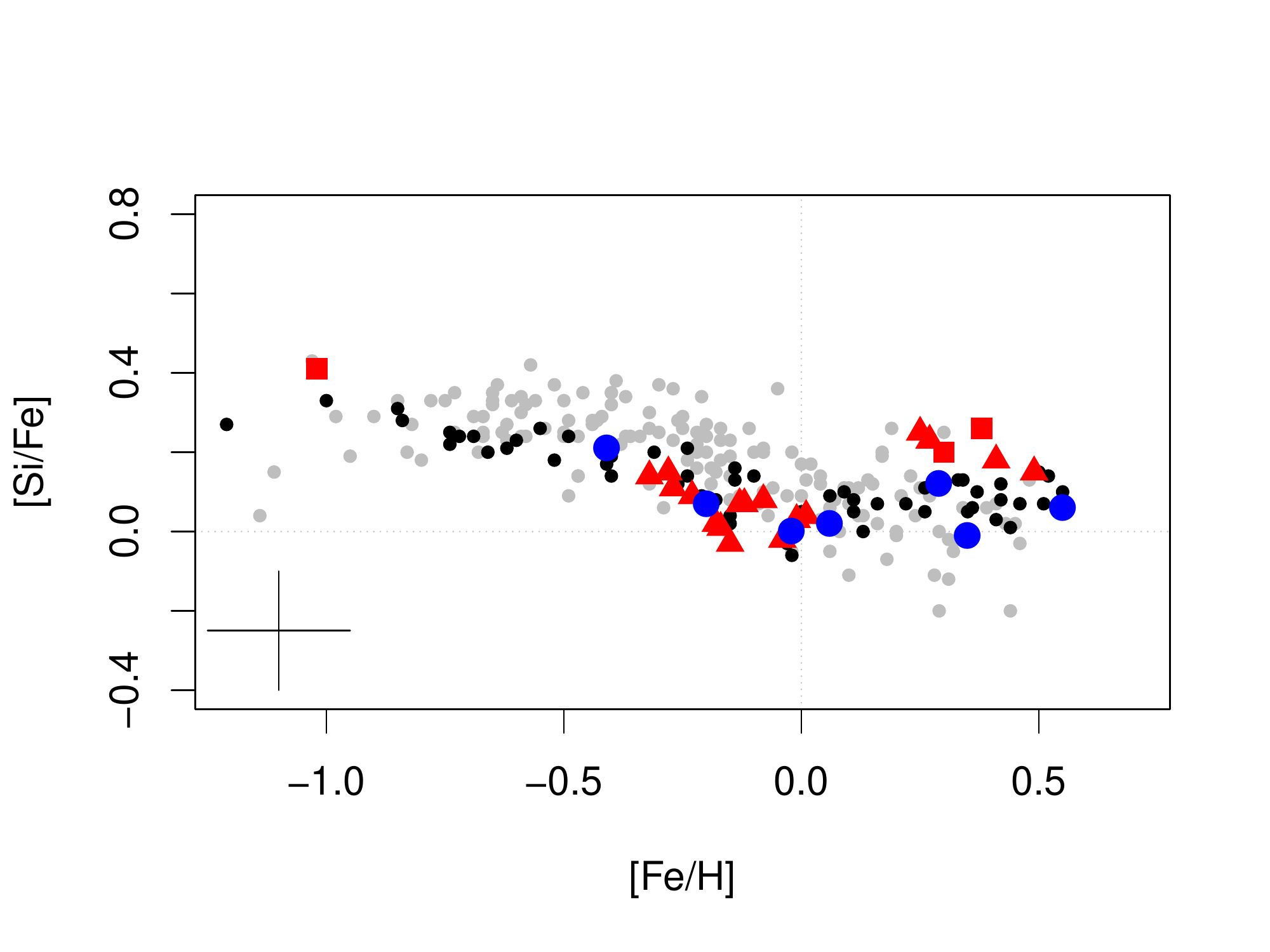} 
\caption{[Si/Fe] vs. [Fe/H]. Red triangles and squares are Nuclear Star Cluster respective Nuclear Stellar Disk stars, blue disks the Milky Way Disk stars, gray dots bulge giants from \citet{johnson:2014}, black dots microlensed bulge dwarfs from \citet{bensby:13}.} \label{fig:sife_bulge}
\end{figure}

We have also reported the first trend of [Si/Fe] vs. [Fe/H] for giants in the NSC and its vicinity.  Below Solar metallicity, we largely confirm the initial study of similar giants in the nuclear disk by \citet{ryde_schultheis:15}.  However, our new results show for the first time a {\it jump} in [Si/Fe] to +0.2\,dex, for [Fe/H]$\,>0$. This particular trend has not been observed in any other stellar population.
\myedit{We emphasize that the sample size is small and the uncertainties of the abundances for Si and Fe in quadrature is $\sim0.2$\,dex, so it is clear that a} confirmation using a larger sample, and additional \textalpha-element lines \myedit{is an essential next step}.
In Figure~\ref{fig:sife_bulge} we compare the [Si/Fe] abundance ratio as a function of [Fe/H] distribution of our samples of NSC (red triangles), NSD (red squares), and MWD (blue triangles) stars with those of giants (gray dots) \citep{johnson:2014} and microlensed dwarfs (black dots) in the bulge \citep{bensby:13}. We find that that our NSC and MWD samples fit the lower envelope of the bulge star distribution up to about solar [Fe/H]. The supersolar stars in the NSC and NSD show enhanced Si abundance compared to our disk and bulge stars.

It is also interesting to compare the properties of the NSC with those of the complex stellar system Terzan 5 in the bulge, which has multiple populations spanning 1.3\,dex in metallicity \citep[e.g.][]{origlia:11,origlia13}. Of special note are the two populations at [Fe/H]=$-0.3$\,dex and [Fe/H]=+0.3\,dex; these two populations have [Si/Fe]=+0.34 and [Si/Fe]=+0.03 respectively, with some stars in the metal rich group have subsolar [Si/Fe]. The ages of these populations have been constrained by \citet{ferraro16} and \citet{origlia:19}. The subsolar component with \textalpha-enhancement was dated at 12\,Gyr, while a younger turnoff of age 4.5\,Gyr appears to be responsible for the metal rich component with about solar-scaled alpha. The trend in \afe\ vs.\ [Fe/H] appears reversed in the NSC compared to Terzan 5, with the \textalpha-enhancement rising at high metallicity, but other \textalpha-elements need to be measured before drawing firm conclusions. The contrast of these two systems is of great interest, and illustrates how chemical evolution can play out differently, even in systems of very similar metallicity.  


\subsection{Chemical Enrichment Scenarios}

Assuming that the enhancement in Si is also present in other \textalpha-elements, the origin of the NSC populations is even more intriguing and we can speculate on possible formation and chemical enrichment scenarios. Some of the key questions for this investigation is ages of stars and the source of gas for star formation.

As per the discussion on Terzan 5 above, the subsolar and supersolar metallicities stars of the NSC can be seen as two groups of stars, and it would be interesting to ponder if there is a clear age dichotomy between them. Unfortunately, isochrone dating of the NSC is not possible because currently we can only observe bright giants near the tip.

We use chemical evolution models \citep{matteucci:12,grieco:15} to explore the chemical enrichment history of the NSC as traced by the sub-solar metallicity stars with about solar-scaled alphas and super-solar metallicity stars with enhanced \textalpha-elements. The models assume that at the very early epoch of the galaxy and bulge formation a star formation episode occurred with a punctual (and probably local) high star formation rates, i.e.\ a mini-starburst activity. The type-II supernovae formed during such a burst would then inject \textalpha-elements within a few Myr. The depth of the gravitational potential well of the Galactic center could help retain or rapidly re-accrete this enriched gas \citep{emsellem:15}, and possibly form the population presented above, before the outbreak of type-Ia supernovae (that would lower the \afe\ content of the interstellar medium).  
This early and massive star forming episode could be followed by a recent starburst occurred a few Gyr ago \citep[][Schödel et al. 2020 in prep.]{pfuhl:11,nogueraslara:19}, but the origin of the gas and its retention in the NSC remain a challenge.

\begin{figure}
\centering
\includegraphics[trim={0cm 0cm 0cm 1.0cm},clip,width=1.00\hsize]{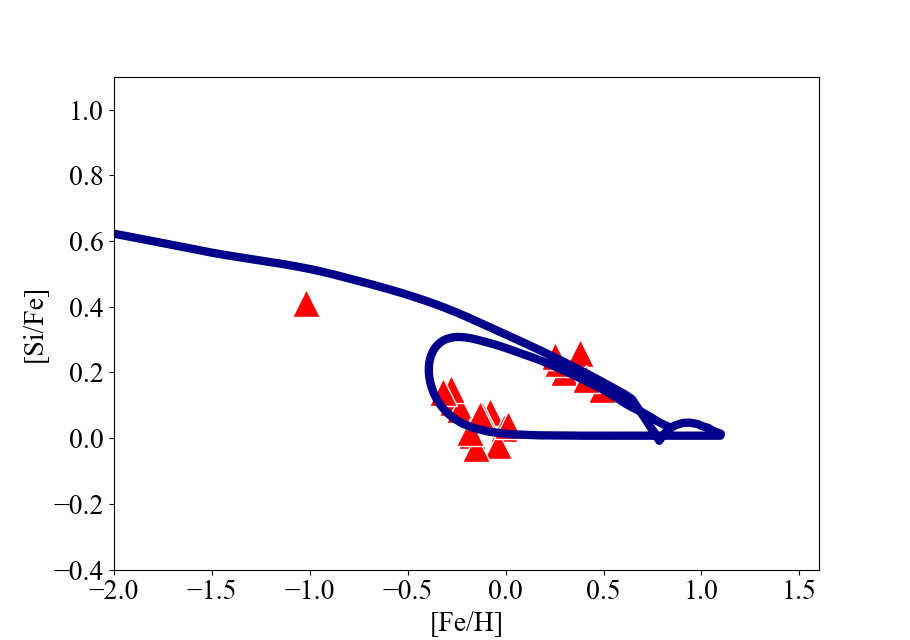} 
\caption{Chemical evolution model with a sudden accretion of primordial gas 2.5\,Gyr ago adopting a \citet{salpeter1955} IMF. The blue line shows the \sife\ vs.\ \feh\ trend of the gas from which stars are formed. Simplified, the line follows a time line from the upper left, down towards the bottom right as gas is consumed by star formation and enriched by Type II SNe and Type Ia SNe. A sudden infall of primordial or slightly enriched gas dilutes the gas resulting in a horizontal line towards the left. As the density of the gas increases the star formation incerases, for instance in a burst if the gas infall is rapid enough, at which an enrichment of Si by Type II SNe occurs until the time when the burst is diluted by the Type I SNe. \myedit{During periods of low density few or no stars are formed, and therefore we do not expect observe stars uniformly populating the track.} \label{fig:primordial}}
\end{figure}

We first explore a model where a starburst is initiated by a sudden accretion of gas, either primordial or slightly enriched, which results in a dilution of chemical abundances (e.g.\ Fe). The model is illustrated in Figure~\ref{fig:primordial}, where the dilution of chemical abundances appears on the plot as a horizontal line followed by an increase of the \afe\ ratio. This is due to the contribution of core-collapse SNe II in the starburst, followed by a decrease due to the Fe production from Type Ia SNe. This behavior appears as a loop, as one can see in Figure \ref{fig:primordial}. This kind of behavior has been shown also by \citet{johnson_weinberg:19} studying starbursts and by \citet{spitoni2019} and \citet{calura:09} in connection with the Galactic disk. In this scenario the super-solar NSC stars could be either old or as young as the sub-solar stars.

To trigger a recent starburst episode, gas needs to be accreted either from other regions of the Galaxy or from an extragalactic origin. In the former case, torques from the bar drive gas inflows toward the center. It is however not clear why such inflow should be intermittent (as required to explain a sudden burst). Simulations of the Milky Way show that discrete accretion events could be due to the fragmentation of the inflowing gas from the Central Molecular Zone (CMZ) at the inner Lindblad resonance \citep{renaud:13, krumholz:15}. The other possibility is the accretion of gas from extragalactic origin, for instance from a satellite galaxy tidally stripped during a passage near the Milky Way's center, i.e.\ on a rather radial trajectory. The gas stripped from a low-mass galaxy would have very low metallicity, whereas the gas from the CMZ could be near or above solar metallicity \citep{morris:96}.


\begin{figure}
\centering
\includegraphics[trim={0cm 0cm 0cm 2.2cm},clip,width=1.00\hsize]{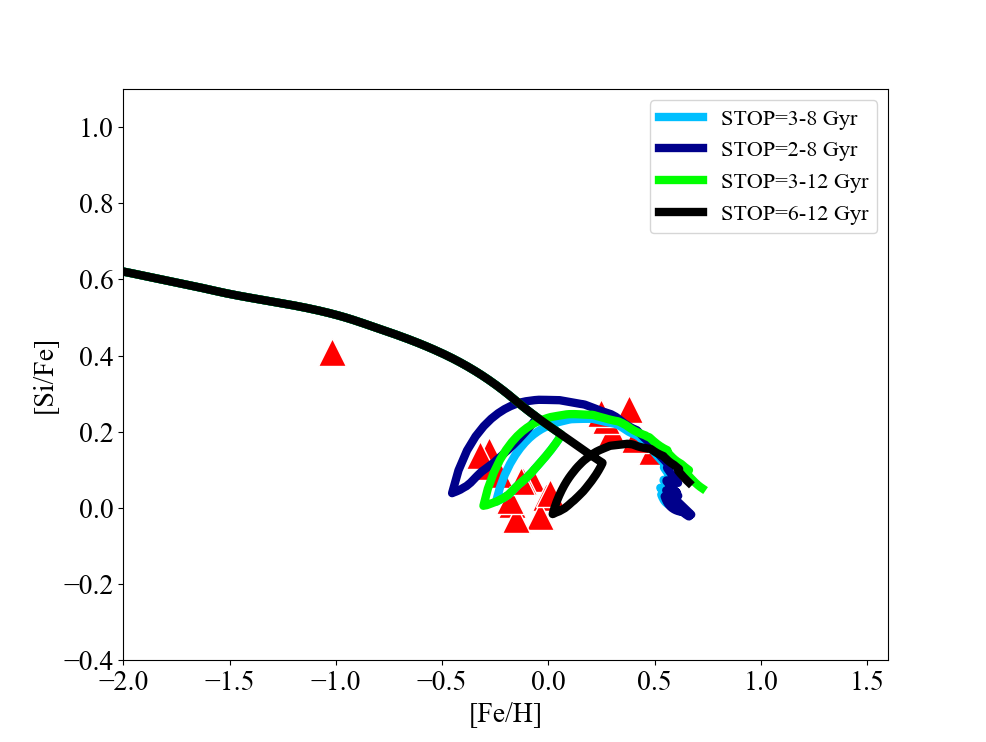} 
\caption{Chemical evolution models varying the temporal extent  of the stop in star formation with a \citet{salpeter1955} IMF. All the cases are characterized by a re-ignition of the  SF after the stop. \myedit{See the description in Figure~\ref{fig:primordial} for how to interpret the line.}
\label{fig:salpeter}}
\end{figure}

Alternatively, we explore a different chemical evolution model where we suppose a pause in the star formation  (see Figure~\ref{fig:salpeter} where pauses of several lengths are assumed). In these cases, no extra infall event is assumed, but simply that the star formation starts again after the stop. As one can see from Figure~\ref{fig:salpeter}, in this case  there is also a loop but not a horizontal decrease of \feh. Here, the loop results from \textalpha-elements decreasing during the pause as Fe continues to be produced by Type Ia SNe, although less than in star formation regime. When star formation reignites the \afe\ ratio increases and then decreases, as explained above.

The main differences between the two scenarios is the extra infall (starburst) and the efficiency of star formation. In the burst case, there is extra infall and the efficiency is high both before and during the starburst, whereas in the case of the pause in the star formation the efficiency is lower especially at the beginning but high after the reignition. The reason is that with a high star formation efficiency, as in some models of the classical bulge/bar \citep{matteucci:19}, the  gas is consumed quickly leaving too little to  trigger star formation after the pause (a similar behavior was suggested by \citealt{matteucci:19} for the bulge). In the case of the bulge it was suggested that buckling of the bar could be the agent responsible for the pause in star formation.

Further investigations into modeling the alpha enrichment of super solar metallicity stars in the NSC and vicinity is called for, and we will investigate this in an upcoming paper.

\section{Conclusions}

We report the first trend of [Si/Fe] vs.\ [Fe/H] for giants in the Nuclear Star Cluster and vicinity. The new trend differs clearly from the local disk; giants with supersolar metallicity show a +0.2\,dex enhancement in [Si/Fe] relative to the stars in the local disk. Such an enhancement is also exceeding the values measured in the bulge. 

In agreement with many studies of the inner and outer bulge, we find the most metal rich stars in the NSC to reach [Fe/H]=+0.5, and exhibit no evidence of more extreme composition. 

\myedit{We emphasize that our current sample size is modest, and it is now} of critical importance to increase the sample size, and to seek confirmation of the trend using other alpha elements. This will allow us to
perform a detailed analysis of the origin and  evolution of the NSC populations by exploring different chemical enrichment scenarios.

\acknowledgments 
We thank Anish Amarsi for hosting B.T. in Heidelberg and assisting with the NLTE calculations. We also thank Mattia Sormani for his comments and suggestions.
B.T.\ and N.R.\ acknowledges support from the Swedish Research Council, VR (project number 621-2014-5640), Funds from Kungl. Fysiografiska Sällskapet i Lund. (Stiftelsen Walter Gyllenbergs fond and Märta och Erik Holmbergs donation), and from the project grant “The New Milky” from the Knut and Alice Wallenberg foundation. M.S.\ acknowledges the Programme National de Cosmologie et Galaxies (PNCG) of CNRS/INSU, France, for financial support. R.M.R.\ acknowledges financial support from his late father Jay Baum Rich. F.R.\ acknowledges support from the Knut and Alice Wallenberg Foundation. A portion of this work was done at the Sexten Center for Astrophysics; the authors acknowledge their hospitality.

The data presented herein were obtained at the W.\ M.\ Keck Observatory, which is operated as a scientific partnership among the California Institute of Technology, the University of California and the National Aeronautics and Space Administration. The Observatory was made possible by the generous financial support of the W.\ M.\ Keck Foundation. The authors wish to recognize and acknowledge the very significant cultural role and reverence that the summit of Mauna Kea has always had within the indigenous Hawaiian community.  We are most fortunate to have the opportunity to conduct observations from this mountain.

\facilities{KECK:II (NIRSPEC)}

\software{SME \citep{sme,sme_code}, REDSPEC \citep{nirspec_reduction}, IRAF \citep{IRAF}, BSYN \& EQWIDTH \citep{marcs:08}}


\bibliographystyle{yahapj}
\bibliography{references}

\end{document}